\documentstyle[prl,aps,multicol,epsfig]{revtex}
\begin{document}
\draft

\title{Non-damped Acoustic Plasmon and Superconductivity in Single Wall Carbon Nanotubes}

\author{Zheng-Mao Sheng$^{1,3}$
\thanks{Email address: zmsheng@mail.hz.zj.cn} and Guoxiang Huang$^{2,3}$}
\address{ $^1$Department of Physics, Zhejiang University, Hangzhou 310028, China\\
          $^2$Department of Physics and Key Laboratory for Optical and Magnetic
              Resonance Spectroscopy, \\
	      East China Normal University, Shanghai 200062,
              China\\
          $^3$International Centre for Theoretical Physics, P.O. Box 586, 
              I-34014 Trieste, Italy         
        }


\maketitle

\begin{abstract}

We show that non-damped acoustic plasmons exist in single wall carbon nanotubes
(SWCNT) and propose that the non-damped acoustic plasmons may mediate 
electron-electron attraction and result in superconductivity in the 
SWCNT. The superconducting transition temperature $T_c$ for the SWCNT (3,3) 
obtained by this mechanism agrees with the recent experimental result 
(Z. K. Tang {\it et al}, {\it Science} {\bf 292}, 2462(2001)\,). 
We also show that it is possible to get higher $T_c$ up to 99 K by doping the SWCNT (5,5).

\end{abstract}

\begin{multicols}{2}

As a novel and potential  carbon material, carbon nanotubes(CNTs) have received a great deal of attention, and 
many of their unique properties have been found\cite{Tanaka99} since they were first discovered \cite{Iijima91}. 
The possibility of superconductivity in CNTs has been discussed based on BCS theory 
(phonon-mediated) \cite{Huang96,Benedict95}. The theoretical prediction given in Ref.\ \cite{Huang96}
about the superconducting transition temperature $T_c$ for single wall carbon nanotubes (SWCNTs) (5,5) with 
diameter of 7\, $\AA$ is only about $1.5 \times 10^{-4}$\,K, and 
it was argued that the electron-phonon interaction is much more favorable to the Peierls transition than 
the BCS superconducting instability. However, a recent experiment on the superconductivity in the smallest SWCNTs
\cite{Iijima00,Wang00} showed that $T_c$ can be up to of 15 K \cite{Tang01}.
This important experimental 
finding requires us to explore a new mechanism for the superconductivity  in the SWCNTs. In this Letter we 
suggest that  the non-Landau damped acoustic plasmons (NDAPs) is responsible for the appearance of 
the superconductivity in the SWCNTs.

The plasmon mediated mechanism for High-$T_c$ superconductivity has attracted much attention 
\cite{Cui91,Ishii93,Sarma98,Semenov00}. Compared to the original BCS phonon 
mechanism, a two dimensional (2D) plasmon with high frequency is favored by the ion-electron mass ratio $M/m^*$, 
as well as 
by a greater energy range, to mediate the electron-electron attraction. The possibility of 
high-$T_c$ superconductivity in the thin wires and quasi-1D organic materials based on the NDAP
\cite{Lee83,Sheng90} mechanism was predicted ten years ago\cite{Lee89,Sheng91}.
The acoustic plasmons in cylindrical quantum-well wires and in CNTs have also been discussed in 
Refs.\ \cite{Wendler94,Lin97,Lin00}.

The acoustic plasmon is the resonant excitation between two kinds of different charge carrier\cite{Ruvalds81}.
The smallness of the radius of SWCNT gives rise to widely separation of transverse single-particle energy 
levels. An electron in particular transverse energy level $E_n$ will have a corresponding effective longitudinal 
Fermi momentum $\hbar k_n = \sqrt{2 m (E_F - E_{n})}$ within the total energy limited by the Fermi energy $E_F$,
where $m$ is the mass of electron. 
The number of the effective longitudinal Fermi levels, corresponding to the number of kinds of the conductive
electrons, is equal to the number of transverse energy levels below Fermi energy. A longitudinal disturbance 
with wavevector $q$ will transfer a momentum $\hbar q$ to each electron. Only those 
electrons within $\hbar q$ neighborhood of the 1D effective Fermi surface $k_n$ are allowed to have a real 
transition, with each of them gaining an energy between $\frac{\hbar^2 q }{m} (k_n-\frac{q}{2})$ and 
$\frac{\hbar^2 q }{m} (k_n + \frac{q}{2})$. In the limit of $q \ll k_{n}-k_{n+1}$, if the disturbance has a 
frequency $\omega \sim \hbar q k_n/m$, all these electrons within $\hbar q$ neighborhood of the 1D effective 
Fermi surface $k_n$ will be resonantly excited. If we denote the polarizability of system by 
$\chi^{(0)}(q,\omega)$, the charge 
density induced by perturbing external potential $\phi_{ex}(q,\omega)$ is
$$ \rho_{in}(q,\omega)= -q^2 \chi^{(0)}(q,\omega)\phi_{ex}(q,\omega)/ \varepsilon(q,\omega).$$  
The resonance is determined by zero of the dielectric function
$\varepsilon(q,\omega)= 1 + 4\pi\chi^{(0)}(q,\omega)$. The mode 
is obviously dominated by the interplay of the transverse energy levels labeled by $n$ and $n+1$. The effective 
perturbing field will cause a parallel polarization in the longitudinally oscillating particles with a 
higher characteristic frequency $\frac{\hbar q }{m} (k_{n}-\frac{q}{2})$ at the $n$ Fermi surface 
but antiparallel polarization in those with a lower characteristic frequency 
$\frac{\hbar q }{m} (k_{n+1}+ \frac{q}{2})$  at the ($n+1$) Fermi surface.
Thus the oscillating particles in level $n$ and $n+1$ will vibrate against each other in the $n$th resonating 
mode, resulting in the formation of NDAPs. 
The wider of separation of transverse levels, the more electron within the neighborhood of the effective 1D 
Fermi surface participate in resonant excitation and result in the higher frequency of the NDAP. Thus the SWCNTs,
being a 1D system, is a good candidate for  existence of the NDAPs.

Because the frequency of the NDAPs is much higher than the phonon frequency and it is non Landau damped, 
it will be more suitable to mediate e-e attraction.  In this Letter we introduce a simple potential 
to model the quasi-1D structure of the SWCNTs, and apply the random phase approximation (RPA) and the 
linear response theory to show the existence of NDAPs in SWCNT(3,3). Then we propose that the NDAPs may mediate 
e-e attraction and result in the superconductivity in the SWCNTs. The superconducting 
transition temperature $T_c$ for SWCNT (3,3) based on this mechanism is calculated and compared with 
the recent experimental result\cite{Tang01}. The possibility for obtaining high $T_c$ for doped SWCNTs 
is also suggested.

The model used is as follows. The electrons are totally confined in an effective potential 
with cylindrical surface of radius $\rho_0$ of a SWCNT and are free to move along the axis of the SWCNT
which we assume to be the $z$ axis.
The single particle Schr\"{o}dinger equation reads

\begin{equation}
 -\frac{\hbar^2}{2m} \left[ \frac{\partial^2}{\partial z^2} +\frac{1}{\rho_0^2}\frac{\partial^2}{\partial 
\varphi^2} \right] \Psi(\rho_0, \varphi, z) = E \Psi(\rho_0, \varphi, z), 
 \label{SCH}
\end{equation}
where $(\rho_0, \varphi, z)$ are cylindrical coordinates. The corresponding energy eigenvalue is given by
\begin{equation}
E_n(k_z)=\frac{\hbar^2 k_z^2}{2 m} + E_n,  \label{Ek}
\end{equation}
where $E_n = \frac{n^2 \hbar^2}{2 m \rho_0^2}$ is the transverse energy level, $n$ is an integer.

We now consider the NDAPs as in Refs.\cite{Lee83,Sheng90}. The Hamiltonian of 
the electron gas confined on the surface of cylinder is 
\begin{equation}
H = - \sum_{j=1} \frac{\hbar^2 \nabla_j^2}{2 m} 
+ \frac{1}{2} \sum_{j \neq k} \frac{e^2}{\vert {\bf r}_j - {\bf r}_k\vert} 
+ H_{+}, 
 \label{Ham}
\end{equation}
where the second term is the Coulomb interaction between electrons and the last term is the contribution from the 
uniformly distributed positive charge background.
By the RPA and the linear response theory, we get a dielectric function as follows\cite{Sheng90}
\begin{eqnarray}
 \varepsilon(q,\omega) & = & 1 + v(q) \sum_{K} u_k[(E_{k+q}-E_{k} -(\hbar \omega + i\eta))^{-1} 
\nonumber \\
&+& (E_{k+q}-E_{k} 
+(\hbar \omega + i\eta))^{-1}], 
 \label{EF}
\end{eqnarray}
where $\eta = 0^{+} $, and 
\begin{eqnarray}
 v(q) &=& \frac{e^2}{2 \pi L}\int_{0}^{2 \pi} d\phi \int_{-\infty}^{+ \infty} dz \frac{e^{i q z}}{\sqrt{z^2 
 + 2 \rho_0^2 (1-cos\phi) }}
 \nonumber \\
&=& \frac{2 e^2}{L} K_0(\sqrt{2} \rho_0 q)
\label{Vq}
\end{eqnarray}
where $L$ is the length of the tube and $K_0(x)$ is the second class zero-th Bessel function.
The form of this $\varepsilon(q, \omega)$ is the same as in 3D system except the difference  in $v(q)$.

 Under zero temperature approximation, we obtain that 
$
\varepsilon(q,\omega) = \varepsilon_1 + i \varepsilon_2
$
with a real part $$ 
 \varepsilon_1 = 1 
+ \frac{2 K_0(\sqrt{2} \rho_0 q)}{\pi q a_0} \sum_{E_n < E_F} \ln \Big\vert\frac{s^2 - {u_n^{+}}^2}
{ s^2 - {u_n^{-}}^2}\Big\vert, 
$$ and an imaginary part 
\begin{displaymath}
\varepsilon_2 = \left\{ \begin{array}{ll}
\frac{2 K_0(\sqrt{2}\rho_0 q)}{\pi q a_0}\sum 1 &\textrm{ if $s$ lies between $u_n^{+}$ and $u_n^{-}$ }\\
0 & \textrm{ otherwise }, \end{array} \right.
\end{displaymath}  
where  $ s =\omega/q $,\, $ a_0 $ is the Bohr radius and $u_n^{\pm} = u_n \pm \frac{\hbar q}{2 m}$, 
$u_n= [2(E_F-E_n)/m]^{1/2}$.

Obviously, $\varepsilon_1 \to \pm\infty$, 
 as $ s \to u_n^{\mp}$. Here $ s $ lies between $  u_n^{-}$ and  $ u_n^{+}$ or between $  u_{n+1}^{+}$ and  
 $ u_n^{-}$. In each case, there always exists a point $\varepsilon_1 = 0 $ between $\varepsilon_1 = - \infty $ 
 and $\varepsilon_1 = + \infty $.  In the former case, $ \varepsilon_2 \neq 0$ which corresponds to a damped 
 mode; but in the second case, the mode $ \varepsilon_2 = 0$ is undamped. As $q$ becomes sufficiently large 
 the damping regions overlap, so all the acoustic plasmon will be damped, and there exists a maximum: 
 $ q_{max} = m(u_n - u_{n+1})/{\hbar} $. The condition for the existence of NDAP is $ \hbar q/m \ll u_n - u_{n+1}
 $.
This means that there are two or more transverse level below the Fermi energy of the system which must be 
sufficiently widely separated so that each level can house a large number of electrons with different $ k_z $, 
leading to the collective nature of the NDAP.

\begin{figure}
\epsfxsize=3.0in \epsfysize=2.0in \epsfbox{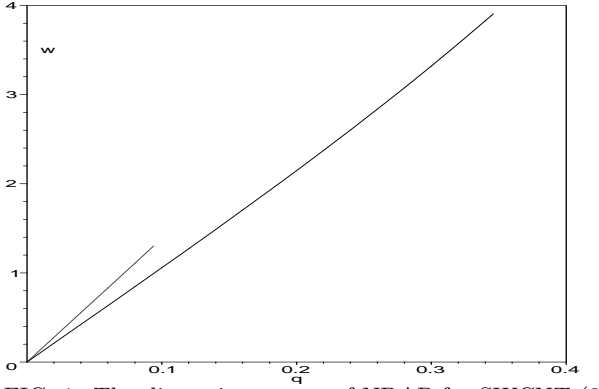}
\caption{The dispersion curves of NDAP for SWCNT (3,3) with radius of 2.1 $\AA$, transverse energy levels $E_0=0$,
$E_1=0.858 $\, ev, $E_2 =3.43 $\, ev and $E_F = 5.88 $\, ev, where the units of $\omega$ and $q$ 
are $10^{15}$\, Hz  and $10^{10}$ m$^{-1}$ respectively. }
\label{fig1}
\end{figure}

Substituting the radius of armchair SWCNT (3,3), $\rho_0 = 2.1 \AA $, into expression of $E_n$, we get $ E_0 = 0$,
$E_1=0.858$\, ev, $E_2=3.43 $\, ev, $E_3 =7.72$\, ev. For a 1D system, if we denote the line density of the electrons 
in the system by $n_1$, then $E_F = \frac{\pi^2 \hbar^2}{8 m} n_1^2$. If there are N electrons per cell, then 
$n_1 = N/a$, for SWCNT (3,3), $a=2.52 \AA$. Taken $N = 2 $, then we get that $E_F = 5.88$\, ev. This means there 
are three transverse energy levels below the Fermi energy. Let $\varepsilon_1$ and 
$ \varepsilon_2$ equal to zero, we can get the dispersion curves of NDAP as shown Fig.\ \ref{fig1}.
From Fig.\ \ref{fig1}  we obtain the $\omega_{NDAP} = 3.9 \times 10^{15}$\, Hz, which is much higher than 
the frequency of phonon.

  Next we cosider the superconductivity in the SWCNTs. Because the property of the NDAPs shown above, we assume 
that e-e attraction may be realized by exchanging 
the NDAPs, leading to electrons pairing. We introduce the following 
Hamiltonian 
\begin{equation}
H = H_1 + H_2 + H_3 
\label{Vq2}
\end{equation}
where $ H_1 = \sum_{{\bf k}, \sigma} E_{{\bf k}} C_{{\bf k}, \sigma}^{+} C_{{\bf k}, \sigma} $ is the kinetic
energy of electrons,
\begin{displaymath}
H_2 = -\frac{1}{2}\sum_{{\bf q}, {\bf k}_1,{\bf k}_2, \sigma}V_{NDAP}C_{{\bf k}_1+{\bf q}, 
\sigma_1}^{+} C_{{\bf k}_2-{\bf q}, \sigma_2}^{+} C_{{\bf k}_2, \sigma_2} C_{{\bf k}_1, \sigma_1}
\end{displaymath} 
is the e-e attraction by exchanging the NDAPs with 
\begin{displaymath}
V_{NDAP}  = \left\{ \begin{array}{ll}
{\rm Const}.\, V & \textrm{ in the $2\hbar \omega_{NDAP}$ energy shell} \\
0 & \textrm{ otherwise } \end{array} \right.
\end{displaymath}
and
\begin{displaymath} 
H_3 = \frac{1}{2}\sum_{{\bf q}, {\bf k}_1,{\bf k}_2, \sigma_1,\sigma_2}V^{c}(q)C_{{\bf k}_1\sigma_1}^{+} 
C_{{\bf k}_2, \sigma_2}^{+} C_{{\bf k}_2-{\bf q}, \sigma_2} C_{{\bf k}_1+{\bf q},, \sigma_1}
\end{displaymath}
is the coulomb interaction between electrons with
\begin{displaymath}
V^{c}(q)  = \left\{ \begin{array}{ll}
e^2 c(q)/\varepsilon(q, 0) &\textrm{ if  $E_{{\bf k}_1}$, $E_{{\bf k}_2} < E_F$ } \\
0 & \textrm{ otherwise }, \end{array} \right.
\end{displaymath}  
where $c(q) =  4 \pi \rho_0 K_0(\sqrt{2}\rho_0 q)$ and  
$~\varepsilon(q,0)$ is the static dielectric function.

Following the usual BCS theory\cite{schrieffer83}, we can get the superconducting transition temperature as
\begin{equation}
k_B T_c = 1.13 \hbar \omega_{NDAP} \exp \left(\frac{-1}{\lambda - \mu^{*}}\right), 
\label{TC}
\end{equation}
where $\lambda = g(0)V$, $\mu^{*} = \frac{\mu}{1 +\mu \ln (E_{F}/\hbar\omega_{NDAP})}$,
$ \mu = \frac{g(0)}{\Omega} \int_{0}^{2 k_F} V^{c}(q) \frac{dq}{2 k_F}$, 
$g(0)$ is the electron state density on the Fermi surface.

For 1D systems, $g(0)= \sqrt{ \frac{m L^2}{2\hbar^2 \pi^2}} \frac{1}{E_F}$.
To estimate $T_c$, we need to know the e-e attraction strength. Suppose that the strength is $V_0$  and zero 
in the effective attraction range $l$ and outside respectively. Similar to the calculation for exchanging Pi meson 
between nucleons, we can get
$ cot(\frac{ m (V_0-B) l^2}{\hbar^2})^{1/2} = (\frac{B}{V_0 -B})^{1/2} $. Here 
$B$ is bind energy, it is approximated by gap $\Delta$. If $B \ll V_0$, 
then $V_0 l^2 = \frac{\pi^2 \hbar^2}{4 m}$. The exchange energy is 
$\Delta p \cdot v_F \simeq \hbar\omega_{NDAP}$, $ l\leq \frac{\hbar}{\Delta p}\simeq \frac{v_F}{\omega_{NDAP}}$, 
where $\Delta p$ is the change of electron momentum. 
Finally, we obtain
$
\lambda = g(0)V \simeq \frac{\pi \hbar \omega_{NDAP}}{8 E_F}.
$\, 
For SWCNT(3,3), as shown above one has $\omega_{NDAP}=3.9 \times 10^{15}$\, Hz and hence  we have 
$\lambda = 0.171$ and $\mu^{*} =0.036 $. Substituting them in to Eq.\
(\ref{TC}), we get $T_c = 19$\,K,
which agrees roughly with the experiment result $T_c = 15$\,K reported in Ref.\cite{Tang01}.

\begin{figure}
\epsfxsize=3.0in \epsfysize=2.0in \epsfbox{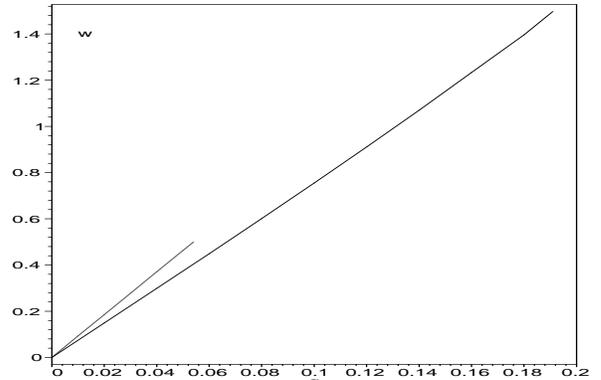}
\caption{ The dispersion curves of NDAP for SWCNT (5,5) with radius of 3.39 $\AA$, 
transverse energy levels $E_0=0$,
$E_1=0.329$\, ev, $E_2 =1.32$\, ev, $E_3=2.96$\, ev and $E_F = 2.61$\, ev. 
The units of $\omega$ and $q$ are $10^{15}$\, Hz  and $10^{10}$ m$^{-1}$ respectively. }
\label{fig2}
\end{figure}

We finally discuss how to get  higher $T_c$ in the SWCNTs. For SWCNT (5,5) with radius of 3.39 $\AA$,
the separation of transverse energy levels is not wide as SWCNT (3,3). 
Its transverse energy levels are $E_0=0$, $E_1=0.329 $\, ev, $E_2 =1.32 $\, ev, $E_3=2.96 $\, ev, $E_4=5.26 $\, ev.
Taking $k_F=.831 \times 10^{10}$ m$^{-1}$ as in Ref.\ \cite{Huang96}, we get  $E_F = 2.61 $\, ev under which 
there are three transverse energy levels. Substituting these parameters into the expression 
of $\varepsilon(q,\omega)$, we can 
obtain the dispersion curves of the NDAPs for SWCNT (5,5) as shown in Fig.\ \ref{fig2} and 
$\omega_{NDAP}=1.5 \times 10^{15} $\, Hz .
Then we have $\lambda =0.148$, $\mu^{*} = 0.039$ and $T_c= 1.3$\, K.
Thus the value of $T_c$ is only $1/15$ of that of SWCNT(3,3). In fact, as the radius of SWCNT increases, the 
separation of 
transverse energy level decreases, so that $\omega_{NDAP}$ becomes lower and leads $T_c$ lower. This conclusion
is in agreement with that obtained in Ref. \cite{Benedict95}. 
However, if one fills the tube with some alkali metallic elements so that the electron can move freely not just 
on the tube surface but in the 
whole cylinder, the single particle Schr\"{o}dinger equation reads
\begin{eqnarray}
& & - \frac{\hbar^2}{2m} \left[ \frac{\partial^2}{\partial z^2} + \frac{1}{\rho} \frac{\partial}{\partial \rho} 
(\rho \frac{\partial}{\partial \rho}) +\frac{1}{\rho^2}\frac{\partial^2}{\partial \varphi^2} \right] 
\Psi(\rho, \varphi, z)
\nonumber \\ 
& &= E \Psi(\rho, \varphi, z). 
 \label{SCH2}
\end{eqnarray}
We can also make a similar calculation as given above. For this situation 
the $V^{c}(q)$ appearing in Eqs.\ (\ref{EF}) and (\ref{Vq}) changes its form. 
Under such condition,  for the doped SWCNT(5,5) with radius of 3.39 $\AA$, the lowest transverse energy 
levels are  $E_{01}= 1.904$\, ev, $E_{11}=4.831$\, ev, $E_{21} = 8.678$\, ev, 
where the subscript of $E_{ n, m}$ are the radial and angular quantum numbers respectively. If a suitable 
conductive electron number density is chosen, there will be   
two or more transverse energy levels under the Fermi level, e.\,g.  $E_F = 5.88$\,ev. 
Substituting this parameters into the expression of $\varepsilon(q,\omega)$, we can obtain the dispersion 
curve of NDAP for doped SWCNT (5,5) and get $\omega_{NDAP}=4.45 \times 10^{15}$\, Hz. 
Then we have $\lambda =0.195$, $\mu^{*} = 0.027$ and hence $T_c= 99$\, K. 
Its $T_c$ is much higher that of undoped SWCNT(5,5). This means it is possible to raise the superconducting 
transition temperature $T_c$ of the SWCNTs by doping  with some alkali metallic elements.

To summarize, we have obtained the dispersion curves of the non damped acoustic plasmons in SWCNTs, 
and proposed a possible new mechanism for superconductivity in these SWCNTs in which 
the electron-electron attraction is mediated by the NDAPs. The smallness of the radius of SWCNT gives rise to 
widely separation of transverse single-particle energy levels.  The wider separation of transverse level, 
the more electrons within the neighborhood 
of the effective 1D Fermi surface participate in resonant excitation and result the higher frequency of NDAPs. 
Moreover, if filling the SWCNTs with alkali metallic elements so that the electron can move freely in the whole 
cylinder, there is much wider separation of transverse level and leads 
much higher frequency of the NDAPs.  However, if the transverse energy level separate too wide so that there is only 
one level under the Fermi surface, it will lose the NDAP.

Because the frequency of the NDAPs is much higher than the frequency of phonon, it is more suitable to mediate 
the e-e attraction. The superconducting transition temperature $T_c =19$\, K for SWCNT(3,3) obtained 
by the NDAP mechanism is in agreement with the experimental result in Ref.\ \cite{Tang01}. We have also pointed 
out that the doped SWCNTs will have much higher transition temperature, if suitable doped materials are chosen
so that there are two transverse energy levels just under the Fermi surface. The value of $\omega_{NDAP}$ 
depends on transverse
energy level structure below Fermi energy, but it is independent of the Fermi energy. However the coupling 
strength of electron pair is
in inverse proportion to $E_F$. So a quasi-1D system which exists two sufficient separated transverse energy 
levels under Fermi energy, but its Fermi energy is not very high, will have higher $T_c$.
For Example, taking $E_F=5.0$\, ev for doped SWCNT(5,5), $\omega_{NDAP}$ is still $4.45 \times 10^{15}$ Hz , 
but $T_c$ is higher than $ 200$\,K.

The authors are grateful for hospitality and support of the Abdus Salam International Centre for 
Theoretical Physics, Trieste, Italy, where this work was carried out. Sheng thanks Yu Lu, T. Xiang, and S.Y. Lou
for useful discussion. 
The work is also supported in part by funds from Pandeng Project of China, NSFC  and Zhejiang Provincial Natural 
Science Foundation of China.

\end{multicols}

\end{document}